\documentclass[aps,pre,notitlepage,nofootinbib]{revtex4-1}

\usepackage{graphicx}
\usepackage{amsmath}
\usepackage{color}

	\usepackage[colorlinks,hyperindex]{hyperref}
 \usepackage{hyperref}
	\hypersetup
	{
		colorlinks,%
		citecolor=blue,%
		linkcolor=blue,%
		urlcolor=black,%
	}

\newcommand{\cH}{{\cal H}}
\newcommand{\cL}{{\cal L}}
\newcommand{\Df}{\Delta_{\rm f}} 
\newcommand{\dn}{{\rm dn}}
\newcommand{\Dw}{\Delta_{\rm w}}
\newcommand{\eg}{{e.g., }}
\newcommand{\gbarw}{\bar\gamma_{\rm w}}
\newcommand{\Gw}{G_{\rm w}}
\newcommand{\gw}{\gamma_{\rm w}}
\newcommand{\half}{\frac{1}{2}}
\newcommand{\hdot}{\dot{h}}
\newcommand{\ie}{{i.e., }}
\newcommand{\kc}{k_{\rm c}}
\newcommand{\pd}{\partial}
\newcommand{\phidot}{\dot{\phi}}
\newcommand{\Pf}{P_{\rm f}}
\newcommand{\Pw}{P_{\rm w}}
\newcommand{\shat}{\hat{s}}

\begin{document}

\title{Pattern transitions in a compressible floating elastic sheet}

\author{Oz Oshri} 
\email{ozzoshri@tau.ac.il} 
\affiliation{Raymond \& Beverly Sackler School of Physics \& Astronomy, Tel Aviv University, Tel Aviv 6997801, Israel}
\thanks{Present address: Department of Chemical and Petroleum Engineering,
University of Pittsburgh, PA 15261, USA}

\author{Haim Diamant} 
\email{hdiamant@tau.ac.il} 
\affiliation{Raymond \& Beverly Sackler School of Chemistry, Tel Aviv University, Tel Aviv 6997801, Israel}

\date{May 16, 2017}

\begin{abstract} 
  Thin rigid sheets floating on a liquid substrate appear, for
  example, in coatings and surfactant monolayers. Upon uniaxial
  compression the sheet undergoes transitions from a compressed
  flat state to a periodic wrinkled pattern to a localized folded
  pattern. The stability of these states is determined by the in-plane
  elasticity of the sheet, its bending rigidity, and the hydrostatics
  of the underlying liquid. Wrinkles and folds, and the
  wrinkle-to-fold transition, were previously studied for
  incompressible sheets. In the present work we extend the theory to
  include finite compressibility. We analyze the details of the
  flat-to-wrinkle transition, the effects of compressibility on
  wrinkling and folding, and the compression field associated with the
  pattern formation. The state diagram of the floating sheet including
  all three states is presented.
\end{abstract}
\maketitle

\section{Introduction}

A thin rigid layer covering a liquid is quite an abundant system. Milk
skin and dense Langmuir monolayers \cite{Lee2008} are two
examples. When a thin sheet is laterally compressed, it
buckles. Unlike the Euler buckling \cite{ElasticityBook} of a sheet of
paper, which makes a single hill or valley, a sheet that covers a soft
substrate buckles in more elaborate shapes. The competition between
the sheet's resistance to bend, favoring a single large buckle, and
the substrate's resistance to deform, preferring many small wiggles,
creates patterns with a characteristic intermediate wavelength
\cite{ThompsonBook,Milner1989,Cerda2003}. The simplest pattern, of
periodic uniaxial waves, {\it wrinkles}, is typically the first to
appear upon lateral compression or confinement
\cite{Milner1989,Cerda2003,Vella2004,Zhang2007,Huang2010,Brau2011,Vanderparre2011,Davidovitch2012,Li2012}. In
this case the deformation is evenly spread across the sheet. In
another pattern, which often develops upon further compression of a
wrinkled sheet, the deformation gets localized in a finite region, a
{\it fold}
\cite{Hunt1993,Lee1996,Pocivavsek2008,Reis2009,Leahy2010,Holmes2010,Diamant2010,Audoly2011,Diamant2011,Brau2013,Diamant2013,Rivetti2013,Semler2013,Demery2014,Rivetti2014,Oshri2015}. The
localization of wrinkles into a fold is intrinsic, \ie it is not tied
to any inhomogeneity in the sheet. The generic patterns of wrinkles
and folds are observed in a large variety of systems
\cite{Vanderparre2011,Schroll2011}, over a wide range of length
scales\,---\,from single-molecule surfactant layers
\cite{Lee2008,Fischer2005,Gopal2006a,Gopal2006b}, to 6-nm-thick
nanoparticle layers \cite{Pocivavsek2008,Leahy2010}, to polymer sheets
with thicknesses of order 10~nm \cite{Huang2007} and 1~$\mu$m 
\cite{Pocivavsek2008,Holmes2010}.

The case of a liquid substrate, whose resistance to deformation arises
from its weight, has attracted particular attention. The fact that the
hydrostatic force on the sheet acts only normal to it, makes this
highly nonlinear problem more tractable than similar ones in
thin-sheet elasticity. For example, the problem of an infinite,
incompressible, floating sheet is exactly solvable
\cite{Diamant2011,Diamant2013,Rivetti2013}, and so is the nonlinear
wrinkling of a finite incompressible sheet \cite{Oshri2015}. The
theory reveals a second-order wrinkle-to-fold transition, and a
narrow stability region for wrinkles, which disappears in the limit of
an infinite sheet. Thus, wrinkles in floating sheets turn out to be a
finite-size phenomenon \cite{Diamant2010}. The theory was found to fit
the experiments to a remarkable accuracy with no fitting parameters
\cite{Brau2013}. (We note that patterns in floating thin sheets can
originate from other effects such as surface tension
\cite{Huang2007,Huang2010,Vella2010,Wagner2011,King2012,Schroll2013,Pineirua2013};
the present work is restricted to substrate effects that are purely
hydrostatic.)

Up until now the theory of floating sheets has been limited to
incompressible layers. Indeed, as the sheet is made increasingly thin,
its resistance to bending becomes increasingly weaker than its
resistance to compression\,---\,while the compression modulus depends
linearly on the thickness $t$, the bending modulus scales as $t^3$
\cite{ElasticityBook}. Hence, the assumption of incompressibility is
valid for a sufficiently small thickness. It does not allow, however,
to examine in detail the onset of wrinkling, \ie the transition from a
compressed flat configuration to a wrinkled one; in the absence of
compressibility, wrinkles must appear at an arbitrarily small
confinement. In addition, it is of interest to study the evolution of
the compression field accompanying the pattern formation, as well as
the corrections to earlier results due to compressibility. These are
the purposes of the present work.

We begin in Sec.~\ref{sec_model} by introducing the model and its
resulting energy functional. Section ~\ref{sec_analysis} presents a
detailed analysis of the energy-minimizing configurations. Following
the footsteps of Refs.~\cite{Audoly2011,Oshri2015}, we employ a
multiple-scale analysis to study the flat-to-wrinkle transition,
wrinkle growth, the wrinkle-to-fold transition, and the subsequent
evolution of the fold. In Sec.~\ref{sec_discuss} we conclude and
discuss the experimental relevance of the findings.

\begin{figure}[tbh]
\vspace{0.7cm}
\centerline{\resizebox{0.6\textwidth}{!}
{\includegraphics{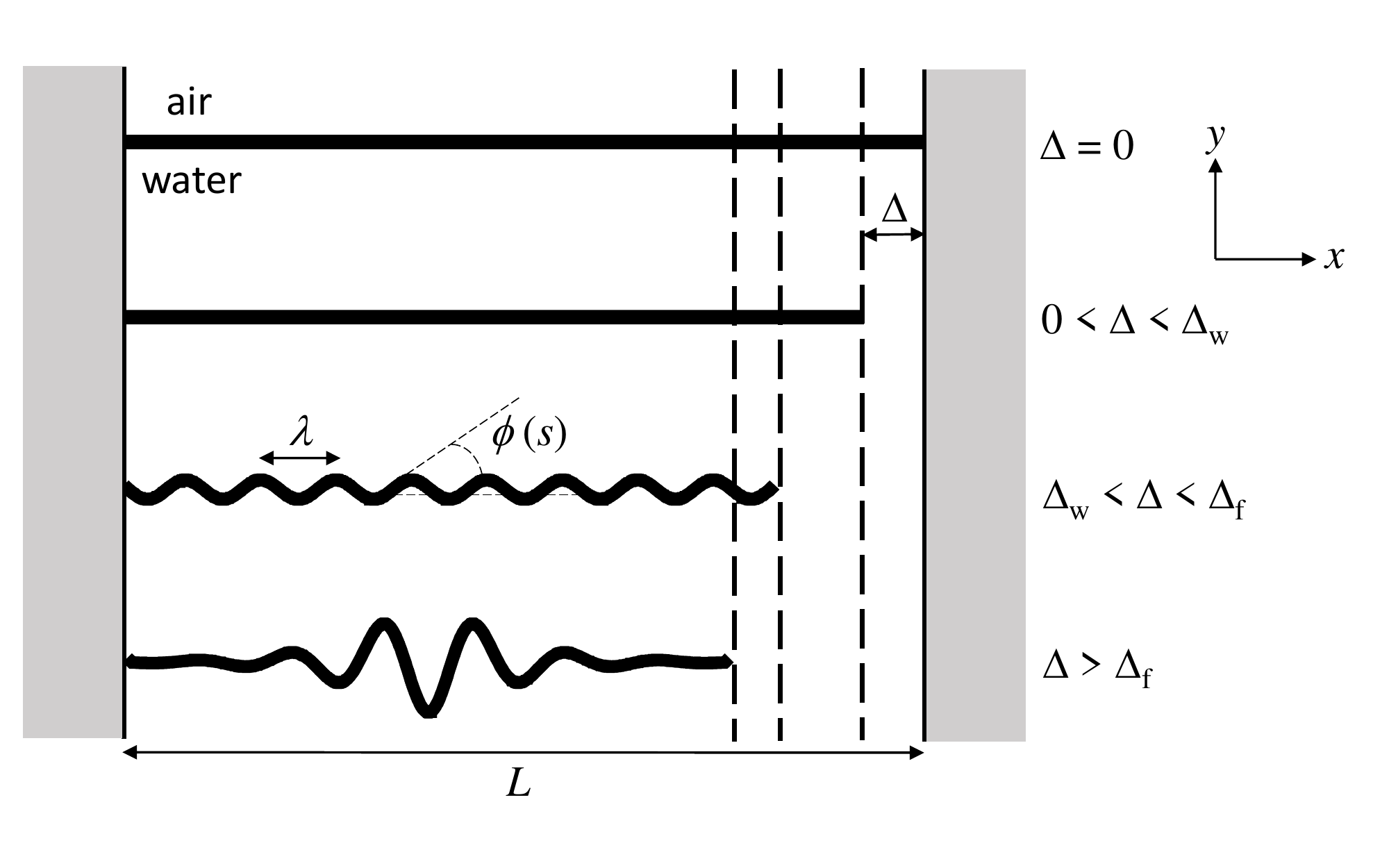}}}
\caption{Schematic evolution of patterns in a compressible finite
  sheet floating on a heavy liquid. For small displacements,
  $\Delta<\Dw$, the sheet is compressed while remaining flat. At a
  critical displacement $\Dw$ the flat sheet buckles into wrinkles of
  wavelength $\lambda$.  For $\Dw<\Delta<\Df$ the wrinkles grow in
  amplitude until, at another critical displacement $\Df$, the pattern
  begins to localize. For $\Delta>\Df$ the pattern continues to
  localize into a single fold.}
\label{fig_scheme}
\end{figure}

\section{Model}
\label{sec_model}

A compressible thin elastic sheet of relaxed length $L$, bending
modulus $B$, and stretching modulus $Y$, lies on a fluid substrate of
mass density $\rho$. The sheet is uniaxially compressed from the
boundaries and develops an out-of-plane deformation on the $xy$ plane,
independent of $z$ (see Fig.~\ref{fig_scheme}). This configuration
makes the problem one-dimensional, dependent solely on the arclength
along the sheet. We distinguish between the relaxed arclength, $s$,
and the compressed one, $\shat$. The deformation of the sheet is
defined by two functions: a compression field, $\gamma(s)=d\shat/ds$
[($\gamma-1$) being the in-plane strain], and the angle profile
$\phi(s)$ between the tangent to the sheet and the horizontal
axis. (The local curvature is then given by $v\equiv
d\phi/d\shat=\gamma^{-1}d\phi/ds$.) It is convenient to define also
the height function $h(s)$, which is related to $\gamma$ and $\phi$ by
the geometric constraint,
\begin{equation}
  \frac{dh}{d\shat} = \gamma^{-1}\frac{dh}{ds} = \sin\phi.
\label{geometric-constraint}
\end{equation}

The energy (per unit length along $z$) of a given deformation contains
contributions from stretching, bending, and the fluid's hydrostatics,
\begin{eqnarray}
  E &=& E_{\rm s} + E_{\rm b} + E_{\rm f}, 
\label{energy}\\
  E_{\rm s} &=& \int_{-L/2}^{L/2} ds \frac{Y}{2} (\gamma-1)^2, \nonumber\\
  E_{\rm b} &=& \int_{-L/2}^{L/2} ds \frac{B}{2} 
  \left( \frac{d\phi}{ds} \right)^2, \nonumber\\
  E_{\rm f} &=& \int_{-L/2}^{L/2} ds \frac{\rho g}{2} h^2\cos\phi. \nonumber
\end{eqnarray}
The boundary displacement associated with a given deformation is
\begin{equation}
  \Delta=\int_{-L/2}^{L/2} ds (1-\gamma\cos\phi).
\label{Delta}
\end{equation}
We note that $E_{\rm s}+E_{\rm b}$ as given in Eq.~(\ref{energy}) is
the known energy functional of the {\it extensible elastica}
\cite{Magnusson2001,Oshri2016}. It can be systematically derived from
the continuous limit of a discrete model, which ensures the
independence of the stretching and bending contributions
\cite{Oshri2016}.  The equilibrium deformation of the sheet is the
minimizer of $E$ for a given $\Delta$, subject to the geometrical
constraint (\ref{geometric-constraint}). Alternatively, instead of
$\Delta$, we can specify the boundary pressure (force per unit length)
$P$ and minimize
\begin{equation}
  G = E - P\Delta,
\end{equation}
for a given $P$.

The problem has two intrinsic lengths (apart from the system length
$L$),
\begin{eqnarray}
  \xi &\equiv& (B/Y)^{1/2},\\
  q^{-1} &=& \lambda/(2\pi) \equiv [B/(\rho g)]^{1/4},
\label{scales}
\end{eqnarray}
which define a dimensionless compressibility parameter,
\begin{equation}
  \zeta \equiv (q\xi)^2 = (B\rho g)^{1/2}/Y.
\end{equation}
For an incompressible sheet, $Y\rightarrow\infty$, and so
$\zeta\rightarrow 0$. For a compressible sheet, in terms of the sheet
thickness $t$, we have $B\sim t^3$ and $Y\sim t$, implying $\xi\sim
t$, $q^{-1}\sim t^{3/4}$, and $\zeta\sim
t^{1/2}\xrightarrow{t\rightarrow 0}0$.  Thus, $\zeta$ is small for
thin sheets. The analysis given in Sec.~\ref{sec_analysis} yields
expressions which are correct to all orders in $\zeta$.

We rescale all lengths by $q^{-1}$, the energies per unit length by $B
q$, and the pressure by $\zeta Y=(B\rho g)^{1/2}$. This results in
\begin{equation}
  G = \int_{-L/2}^{L/2}ds \left[ \frac{1}{2\zeta}(\gamma-1)^2 +
    \half\left(\frac{d\phi}{ds}\right)^2 + \half \gamma h^2\cos\phi
    - P(1-\gamma\cos\phi) \right].
\label{G}
\end{equation}
Thus, the problem contains three dimensionless parameters: $\zeta$ and
the rescaled $P$ and $L$.

\section{Detailed analysis}
\label{sec_analysis}

As in earlier works \cite{Diamant2011,Diamant2013,Oshri2015}, we first
recast the problem in dynamical terms, where $s$ stands for time and
$(\,\dot{ }\,)$ denotes a derivative with respect to $s$. In this
analogy we need to minimize an ``action'', ${\cal S}=\int_{-L/2}^{L/2}
ds {\cal L}(\phi,h,\gamma,\dot\phi,\dot h,\dot\gamma)$, with the
``Lagrangian'',
\begin{equation}
  \cL=\frac{1}{2}\dot{\phi}^2+\frac{1}{2\zeta}(\gamma-1)^2
  +\frac{1}{2}\gamma h^2
  \cos\phi-P(1-\gamma\cos\phi)-Q(s)(\gamma\sin\phi-\dot{h}).
\label{cL}
\end{equation}
In Eq.~(\ref{cL}) $Q(s)$ is a Lagrange multiplier introduced to impose
the local geometrical constraint (\ref{geometric-constraint}) at each
point along the sheet. The conjugate ``momenta'' are identified as
$p_{\phi}=\pd\cL/\pd\phidot=\phidot$, $p_h=\pd\cL/\pd\hdot=Q$, and
$p_{\gamma}=\pd\cL/\pd\dot{\gamma}=0$. These are used to construct the
``Hamiltonian'', $\cH(\phi,h,\gamma,p_\phi,p_h,p_\gamma) =
p_{\phi}\phidot + p_{h}\hdot + p_{\gamma}\dot{\gamma} - \cL$,
\begin{equation} 
  \cH = -\frac{1}{2\zeta}(\gamma-1)^2 + \half p_\phi^2 - \left(
  \half h^2 + P \right)\gamma\cos\phi + p_h\gamma\sin\phi
  = \cH_0 = \mbox{const}.
\label{cH}
\end{equation}
As $\cL$ has no explicit dependence on $s$, $\cH=\cH_0$ is a
``constant of motion'', \ie it is uniform across the sheet and fully
determined by the conditions at the boundaries.

\subsection{General relation between compression and curvature}
\label{sec_relation}

The fact that the energy does not depend on $\dot{\gamma}$ has led
above to $p_\gamma=0$. This results, through Hamilton's equation,
$\dot{p}_{\gamma}=-\pd\cH/\pd\gamma=0$, in
\begin{equation}
  \frac{1}{\zeta}(\gamma-1) + \left(\half h^2+P\right)\cos\phi
  - p_h\sin\phi = 0.
\label{gamma1}
\end{equation}
Equations (\ref{cH}) and (\ref{gamma1}) yield a general relation
between the compression and angle fields,
\begin{equation}
  \gamma(s) = \left[ \bar\gamma^2 -
    \zeta\left(\frac{d\phi}{ds}\right)^2 \right]^{1/2},
\label{gamma}
\end{equation}
where $\bar\gamma=1+2\zeta \cH_0$ is a constant determined by boundary
conditions. 

This relation raises several noteworthy points.  (i) In the
incompressible limit ($\zeta=0$), $\gamma(s)\equiv 1$ as
expected. (ii) In flat configurations, the compression field has a
uniform value, $\gamma(s)\equiv\bar\gamma$, dictated by boundary
conditions. As the sheet goes out of plane, the compression is
increased throughout the sheet, $\gamma(s)<\bar\gamma$, with maxima at
the points of maximum curvature. (In the case of uniaxial deformation,
unlike the case of a two-dimensional one, buckling does not cause
stress collapse.) (iii) If the sheet is hinged at its boundaries
(having zero curvature there), $\bar\gamma$ is the boundary value of
the compression field, and $\gamma(s)$ is smaller than that value
everywhere in the sheet. (iv) The relation (\ref{gamma}) is
independent of the hydrostatic contribution $E_{\rm f}$; the
dependence on $\rho g$ enters through factors of $q^2$ in both $\zeta$
and the rescaled $ds^2$, which mutually cancel. In fact, the same
relation holds for any energy of the form $E_{\rm s}+E_{\rm b} + \int
ds\gamma V(\phi)$, where $V(\phi)$ is an arbitrary function
\cite{Oshri2016}. (v) In terms of the curvature,
$v=d\phi/d\shat=\gamma^{-1}d\phi/ds$, Eq.~(\ref{gamma}) takes the
form,
\begin{equation}
  \gamma(\shat) = \frac{\bar\gamma} {\sqrt{1+\zeta v^2}}.
\end{equation}
This relation between stretching and curvature is reminiscent of
relativistic mechanics, where $v$, $-\zeta^{-1/2}$, and $\gamma$ play
the roles, respectively, of velocity, speed of light, and the Lorentz
dilation factor, such that the incompressible limit corresponds to
classical mechanics. This curious (and useful) analogy is discussed in
Ref.~\cite{Oshri2016}.

\subsection{States and transitions}

In this section we derive expressions for the in-plane deformation,
out-of-plane profile, and critical values of pressure and displacement
at the two transitions. The expressions are valid for an arbitrary
size of the compressibility parameter $\zeta$. However, since this
parameter is typically very small, a leading-order expansion in
$\zeta$ should be sufficient for all practical purposes, and we
provide these approximate expressions as well.

We begin with the flat state. Specializing Eqs.~(\ref{Delta}),
(\ref{G}), and (\ref{gamma}) to the case of $h(s)=\phi(s)\equiv 0$, we
obtain
\begin{eqnarray}
  \gamma(s) &\equiv& \bar\gamma = 1 - \Delta/L = 1 - \zeta P, \nonumber\\
  \Delta &=& L\zeta P, \nonumber\\
  G &=& -L \zeta P^2/2, 
\label{flat}
\end{eqnarray}
which describes the simple linear deformation of a uniaxially
compressed flat sheet.

To analyze the non-planar patterns we employ a multiple-scale analysis
similar to the one presented in Refs.~\cite{Audoly2011,Oshri2015}. We
assume a long (but finite) sheet, $L\gg k^{-1}$, where $2\pi/k \sim
\lambda$ is the wrinkle wavelength, and $\lambda$ is the intrinsic
length defined in Eq.~(\ref{scales}). Our small parameter is the
distance to the flat-to-wrinkle transition; more specifically,
\begin{equation}
  \epsilon \equiv (\Pw-P)^{1/2}. 
\label{epsilon}
\end{equation}
As $L$ is increased to arbitrarily large values, the wrinkle-to-fold
transition occurs arbitrarily close to the flat-to-wrinkle one
\cite{Diamant2010,Audoly2011}. Thus, restricting the analysis to the
leading order in $1/L$ allows the use of the same perturbation theory
for both transitions.

For simplicity, we limit the discussion to $L=\pi N/k$, where $N$ is
an integer, \ie the wrinkled pattern contains $N$ wrinkles. Odd and
even $N$ correspond respectively to symmetric and antisymmetric
solutions. In addition, we assume hinged boundary conditions, \ie
vanishing height and bending moment at the edges of the sheet,
\begin{equation}
  h(\pm L/2) = \phidot(\pm L/2) = 0.
\label{bc}
\end{equation}
For these boundary conditions, according to Eq.~(\ref{gamma}), the
constant $\bar\gamma$ coincides with the boundary
compression $\gamma(\pm L/2)$. 

In the wrinkled state the pattern is periodic, and slightly above the
wrinkle-to-fold transition it is weakly localized. Thus, the system
exhibits either one spatial variation, or two with well-separated
scales\,---\,fast undulations of wavelength $2\pi/k\sim\lambda$, and a
slow envelope of decay length $\eta^{-1}\gg\lambda$. This suggests
the following approximation:
\begin{equation}
  h(s) \simeq
  \epsilon \cos(k s) H(S),
\label{h-expansion}
\end{equation}
where $S\equiv\epsilon s$ is the slow variable, and we will denote a
derivative with respect to $S$ by $(\,)'$. In the wrinkled state $H$
is constant, and in the folded state it is a decaying envelope. We
have chosen a symmetric profile; an antisymmetric one is obtained by
replacing the cosine with a sine and leads to similar results. The
boundary conditions of vanishing height are satisfied by the fast
oscillating function, $\cos(N\pi/2)=0$ for odd $N$. The boundary
conditions of vanishing bending moment turn into
\begin{equation}
  H'(\pm\epsilon L/2) = 0.
\label{bcH}
\end{equation} 

We substitute the height function (\ref{h-expansion}) in the equations
derived above (specifically, Eqs.~(\ref{geometric-constraint}),
(\ref{Delta}), (\ref{G}), and (\ref{gamma})), average over the fast
oscillations, and obtain approximate expressions for the various
quantities,
\begin{eqnarray}
  G &\simeq& \Gw + \epsilon^2 G_2 + \epsilon^4 G_4,\ \ \ \ \ \
  \Gw = -\zeta\Pw/2,
\label{expansion}\\
  \Delta &\simeq& \Dw + \epsilon^2 \Delta_2 + \epsilon^4 \Delta_4, \ \ \ \ \ \
  \Dw = L\zeta\Pw,
 \nonumber\\
  \bar\gamma &\simeq& \gbarw + \epsilon^2 \bar\gamma_2 
  + \epsilon^4 \bar\gamma_4,\ \ \ \ \ \ \ \ \ \gbarw = 1-\zeta\Pw,
 \nonumber\\
  \gamma(s) &\simeq& \gw + \epsilon^2 \gamma_2(s) +
  \epsilon^4 \gamma_4(s), \ \ \gw = \gbarw = 1-\zeta\Pw,
 \nonumber
\end{eqnarray}
where the zeroth-order terms, the ones at the wrinkle-to-fold
transition, have been obtained from the edge of the flat state,
Eq.~(\ref{flat}).  Since the system has an up-down symmetry, only even
powers of $\epsilon$ survive. The assumed height function
(\ref{h-expansion}), linear in $\epsilon$, is sufficient for getting
the leading terms specified in Eq.~(\ref{expansion}). The expansion is
worked out in detail in the Supplementary Material \cite{suppl}.

\subsubsection{Flat-to-wrinkle transition and wrinkle growth}

The expression for the second-order energy term, $G_2$, is found to be
\cite{suppl}
\begin{equation}
  G_2 = (1 - \gw)L + \frac{1}
  {4\gw^2} \left(k^4-\gw\Pw k^2 + \gw^3\right)
  \int_{-L/2}^{L/2}H^2 ds.
\end{equation}
The flat state becomes unstable when going out of plane lowers the
energy, \ie when the coefficient of $\int H^2$ in $G_2$ becomes
negative, which is obtained when the pressure exceeds $\Pw =
(k^4+\gw^3)/(k^2\gw)$.  This occurs first for a critical wavenumber,
$k=\kc=\gw^{3/4}$, whereby $\Pw=2\gw^{1/2}$.  Adding to these results
the known conditions at the edge of the flat state,
Eq.~(\ref{expansion}), and solving for $\Pw$, we obtain the critical
flat-to-wrinkle wavenumber, pressure, displacement, and compression as
functions of the compressibility parameter alone,
\begin{eqnarray}
  \kc(\zeta) &=& \left(\sqrt{1+\zeta^2}-\zeta \right)^{3/2}
  \simeq 1 - 3\zeta/2, \nonumber\\
  \Pw(\zeta) &=& 2\left( \sqrt{1+\zeta^2} - \zeta\right)
  \simeq 2 (1-\zeta),
\nonumber \\
  \Dw(\zeta) &=& 2L\zeta \left( \sqrt{1+\zeta^2} - \zeta\right) \simeq
  2L\zeta.  \nonumber\\
  \gw(\zeta) &=& 1 - 2\zeta\left( \sqrt{1+\zeta^2} - \zeta\right)
  \simeq 1 - 2\zeta.
\label{PW}
\end{eqnarray}
Equation~(\ref{PW}) extends the results for the wrinkling transition
in an incompressible sheet \cite{Milner1989} ($\kc=1$, $\Pw=2$,
$\Dw=0$, $\gw=1$) to sheets of finite compressibility. We find that
the compressibility lowers the critical pressure and increases the
wrinkle wavelength.  In addition, we obtain the second-order
expressions for the displacement and compression \cite{suppl},
\begin{eqnarray}
  \Delta_2 &=& 
  \frac{1}{4} \sqrt{1+\zeta^2} \int_{L/2}^{L/2} H^2 ds - L\zeta, 
\label{Delta2}\\
  \gamma_2(s) &=& 
  \zeta\left[1 - \half\cos^2(\kc s) H^2(S) \right].
\label{gamma2}
\end{eqnarray}
Thus, as confinement is increased and wrinkles grow, both the mean
compression and its undulations increase as well. 

To get the details of the wrinkled state beyond the transition, we
need the height profile that minimizes the energy. This requires the
fourth-order energy term \cite{suppl},
\begin{equation}
  G_4 = -L\zeta/2 + \frac{2}{\kc^{2/3}} \int_{-L/2}^{L/2} \left[
    \half(H')^2 - \frac{\alpha}{4}H^4 + \frac{\beta}{2}H^2 -
    \frac{1}{4} \left(H H'\right)' \right] ds,
\label{G4}
\end{equation}
where
\begin{equation}
  \alpha = \frac{1}{8} \left[ 1 - \frac{5}{4}\zeta
  \left(\sqrt{1+\zeta^2}-\zeta\right) \right],\ \ \ 
  \beta = \frac{1}{4} \left[ 1 - \zeta \left(
  \sqrt{1+\zeta^2}-\zeta \right) \right]
\label{alphabeta}
\end{equation}
are strictly positive functions of $\zeta$.  The last
(total-derivative) term in Eq.~(\ref{G4}) vanishes due to the hinged
boundary conditions (\ref{bcH}). Minimizing $G_4$ with respect to
$H(S)$ yields the following amplitude equation:
\begin{equation}
  H'' + \alpha H^3 - \beta H = 0.
\label{amplitude}
\end{equation}
Equation (\ref{amplitude}) is to be solved together with the boundary
conditions (\ref{bcH}).

One solution is a constant,
\begin{equation}
  H(S) \equiv H_0(\zeta) = \sqrt{\beta/\alpha}
  \simeq \sqrt{2}\,(1 + \zeta/8).
\label{H0}
\end{equation}
This corresponds to periodic wrinkles,
\begin{equation}
  h(s) = A\cos(\kc s),\ \ \ A=\epsilon H_0,
\label{hwrinkle}
\end{equation}
whose amplitude grows with the drop in pressure as
\begin{equation}
  A(P) = H_0(\zeta) (\Pw-P)^{1/2} \simeq \sqrt{2}
  (\Pw-P)^{1/2}(1+\zeta/8).
\label{APW}
\end{equation}
To get the amplitude as a function of displacement we substitute
$H=H_0$ in Eq.~(\ref{Delta2}), finding
\begin{equation}
  A(\Delta) = 2 \left( \frac{\Delta-\Dw}{L} \right)^{1/2} 
  \left( \sqrt{1+\zeta^2} - 4\zeta/H_0^2 \right)^{-1/2}
  \simeq 2 \left(\frac{\Delta-\Dw}{L}\right)^{1/2} 
  (1 + \zeta).
\label{AW}
\end{equation}
This, together with Eq.~(\ref{epsilon}), yields the
pressure-displacement relation,
\begin{equation}
  P(\Delta) = \Pw - \frac{4(\Delta-\Dw)}{L}\,
  \left(H_0^2 \sqrt{1+\zeta^2} - 4\zeta \right)^{-1}
  \simeq \Pw - 2\frac{\Delta-\Dw}{L} (1 + 7\zeta/4).
\label{PDeltaW}
\end{equation}

Equations (\ref{H0})--(\ref{PDeltaW}) generalize the results known for
wrinkles in incompressible sheets
\cite{Diamant2010,Audoly2011,Diamant2011} ($H_0=\sqrt{2}$,
$A=\sqrt{2(\Pw-P)}=2\sqrt{\Delta/L}$, $P=\Pw-2\Delta/L$) to
compressible sheets. Note that the seeming increase of amplitude with
compressibility in Eqs.~(\ref{APW}) and (\ref{AW}) is misleading; once
the dependence of $\Pw$ and $\Dw$ on $\zeta$ is included, the
amplitude at fixed $P$ or $\Delta$ decreases with $\zeta$, as
intuitively expected.

We also calculate the compression field along the sheet by
substituting the wrinkled profile (\ref{hwrinkle}) in
Eq.~(\ref{gamma2}),
\begin{equation}
  \gamma(s) = \gw + \zeta (\Pw-P) \left[ 1 - 
  \half H_0^2 \cos^2(\kc s) \right]
  \simeq \gw + \zeta (\Pw-P) [1 - \cos^2(\kc s)],
\end{equation}
or, in terms of displacement,
\begin{equation}
  \gamma(s) = \gw + \frac{4\zeta(\Delta-\Dw)}{L} 
   \left( H_0^2\sqrt{1+\zeta^2} - 4\zeta \right)^{-1}
  \left( 1 - \half H_0^2 \cos^2(\kc s) \right)
  \simeq \gw + 2\zeta \frac{\Delta-\Dw}{L} [1-\cos^2(\kc s)].
\end{equation}
The undulating compression field has double the frequency of the
wrinkles. Around the wrinkles' extrema (both minima and maxima) there
is some extra compression, which takes some of the imposed
displacement and makes the wrinkle amplitude slightly smaller than it
would have been in the incompressible case.

The wrinkle amplitude $A$ can be defined as the order parameter of the
flat-to-wrinkle transition. Equations (\ref{APW}) and (\ref{AW}) show
that the transition is second-order, as in the incompressible case.

\begin{figure}[tbh]
\vspace{0.7cm}
\centerline{\resizebox{0.48\textwidth}{!}
{\includegraphics{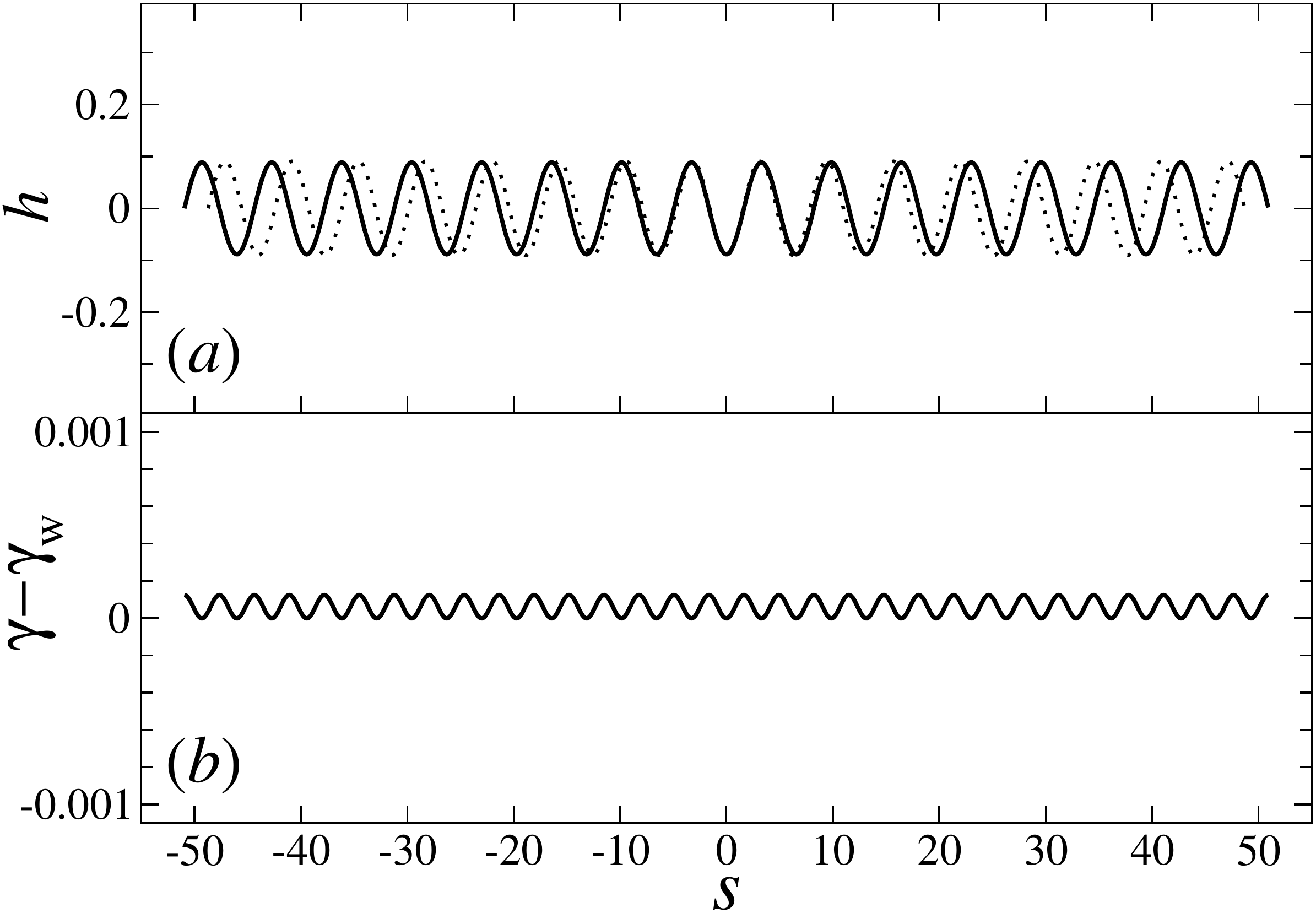}}
\hspace{0.2cm}
\resizebox{0.48\textwidth}{!}
{\includegraphics{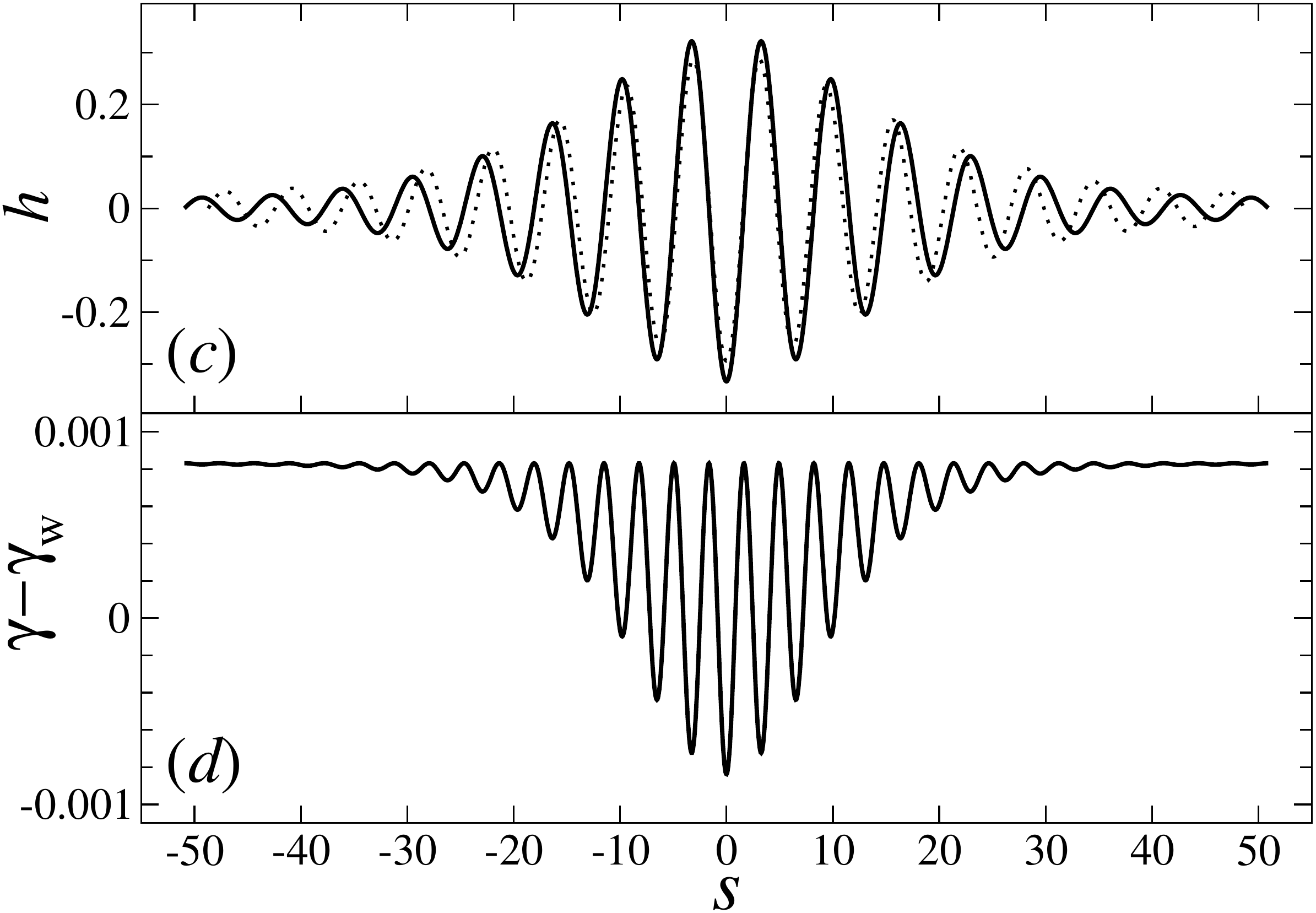}}}
\caption{Height (a,c) and compression (b,d) profiles, as a function of
  arclength, in the wrinkled state (a,b) and folded state (c,d). The
  height profiles of the corresponding incompressible sheet
  ($\zeta=0$; dotted curves) are shown for comparison. The height $h$
  and arclength $s$ are rescaled by the intrinsic length $q^{-1}$ ($q$
  being the wrinkle wavenumber in the incompressible sheet). The
  compression at the flat-to-wrinkle transition, $\gw$, has been
  subtracted from the compression field. The selected parameters are
  $\zeta=0.03$ and $L=31\pi/\kc$, allowing for a symmetric profile
  with exactly 16 wrinkles. These yield $\gw\simeq 0.94$ and critical
  displacements (rescaled by $q^{-1}$) $\Dw\simeq 5.93$, $\Df\simeq
  6.32$. The results are for $\Delta=\Dw+0.2\in(\Dw,\Df)$ (a,b), and
  $\Delta=\Dw+0.6 > \Df$ (c,d). The corresponding values for the
  incompressible sheet (dotted curves) are $\Dw=0$, $\Df\simeq 0.405$,
  and $\Delta=0.2$ (a), $\Delta=0.6$ (c), to ensure a valid
  comparison.}
\label{fig_pattern}
\end{figure}

Figure~\ref{fig_pattern}(a) shows the wrinkled profile of the
compressible sheet along with its incompressible counterpart (dotted
curve). One can notice the slightly shorter wavelength in the
incompressible case. In Fig.~\ref{fig_pattern}(b) we see the
corresponding compression field (minus its uniform value at the
wrinkling threshold).

\subsubsection{Wrinkle-to-fold transition and fold evolution}

Equations (\ref{amplitude}) and (\ref{bcH}) have other solutions in
the form of Jacobi elliptic functions \cite{Abramowitz}. Out of those
twelve functions, only one, $\dn$, is found to provide a physical
solution \cite{Oshri2015},
\begin{eqnarray}
  &&H(S) = \sqrt{\frac{2}{\alpha}} \kappa\, \dn(\kappa S,m),
  \ \ \ \kappa(m) = \sqrt{\frac{\beta}{2-m}},
\label{Hfold}\\
  &&K(m)/\kappa(m) = \epsilon L/2,
\label{Km}
\end{eqnarray}
where $K(m)$ is the complete elliptic integral of the first kind,
which is half the period of the function $\dn$
\cite{Abramowitz}. 
The resulting height profile is
\begin{equation}
  h(s)=\epsilon\sqrt{\frac{2}{\alpha}}\kappa \cos(\kc s)\dn(\kappa
  \epsilon s,m)=\sqrt{\frac{8}{\alpha}}\frac{K(m)}{L}\cos(\kc
  s)\dn\left(\frac{2K(m)}{L}s,m\right).
\label{hfold}
\end{equation}
For given $P$, $L$, and $\zeta$, one first solves Eq.~(\ref{Km}) to
find $m$ and then uses it in Eq.~(\ref{hfold}) to obtain the profile.

The modulus $m$ takes values in the range $0\leq m\leq 1$.  In the
limit $m \rightarrow 0$, the function $\dn(u,m)\rightarrow 1$,
leading to
\begin{equation}
  m\rightarrow 0:\ \ \ h(s) = \epsilon \sqrt{\beta/\alpha} \cos(\kc s),
\end{equation}
which coincides with the wrinkled profile obtained above,
Eq.~(\ref{hwrinkle}). For any $m>0$ the function $\dn(u,m)$ is a
nonuniform envelope, peaked at $u=0$ and decaying symmetrically on both
sides. This breaks the profile's periodicity and localizes it around
the origin. Thus, we can define $m$ as the order parameter of the
wrinkle-to-fold transition. In the limit $m\rightarrow 1$ we have
$\dn(u,m)\rightarrow 1/\cosh(u)$, which gives the following localized
profile:
\begin{equation}
  m\rightarrow 1:\ \ \ h(s) = \epsilon \sqrt{\frac{2\beta}{\alpha}} 
  \frac{\cos(\kc s)}{\cosh\left(\sqrt{\beta}\, \epsilon s\right)}. 
\label{localized}
\end{equation}

Next, we express the pressure and displacement as functions of $m$.
Equation~(\ref{Km}) readily gives
\begin{equation}
  P(m) = \Pw - \frac{4}{\beta L^2}(2-m) K^2(m)
  \simeq \Pw - \frac{16}{L^2}(1 + \zeta)(2-m) K^2(m).
\label{Pm}
\end{equation}
Within our approximations, the maximum height of the localized profile
is proportional to $\epsilon$ even as $m\rightarrow 1$; see
Eq.~(\ref{localized}). Hence, we may continue to use
Eqs.~(\ref{Delta2}) and (\ref{gamma2}) for the displacement and
compression. Substituting in Eq.~(\ref{Delta2}) the envelope $H(S)$ of 
Eq.~(\ref{Hfold}) and using Eq.~(\ref{Km}), we find
\begin{equation}
  \Delta(m) = \Dw + \frac{2}{\beta L} K(m) \left[ 
  H_0^2 \sqrt{1+\zeta^2} E(m) - 2\zeta(2-m) K(m) \right],
\label{Deltam}
\end{equation}
where $E(m)$ is the complete elliptic integral of the second kind
\cite{Abramowitz}.

To obtain the critical wrinkle-to-fold pressure and displacement we
take the limit $m\rightarrow 0$, which yields
\begin{eqnarray}
  \Pf &=& \Pw - \frac{2\pi^2}{L^2\beta} \simeq \Pw - \frac{8\pi^2}{L^2}
  (1 + \zeta), \nonumber\\
  \Df &=& \Dw + \frac{\pi^2}{2\beta L} \left( H_0^2\sqrt{1+\zeta^2}
  -2\zeta \right) 
  \simeq \Dw + \frac{4\pi^2}{L} (1 + \zeta/4).
\label{PfDf}
\end{eqnarray}
These results show that with increasing $L$ the wrinkle-to-fold
transition gets increasingly close to the flat-to-wrinkle one, as
stated earlier. Equation (\ref{PfDf}) extends the results for an
incompressible finite sheet \cite{Oshri2015} ($\Pf=\Pw-8\pi^2/L^2$,
$\Df=4\pi^2/L$). Compressibility pushes the transition to larger
displacement. Expanding $P(m)$ and $\Delta(m)$ in small $m$, we obtain
$\Pf-P \sim \Delta-\Df \sim m^2$. Thus, compressibility does not
affect the order of the wrinkle-to-fold transition, which is
second-order.

Above the wrinkle-to-fold transition, the pressure--displacement
relation is obtained parametrically as $(\Delta(m),P(m))$ from
Eqs.~(\ref{Pm}) and (\ref{Deltam}). In the localized limit,
$m\rightarrow 1$, these equations turn into $P(m) \simeq \Pw -
[4/(\beta L^2)]K^2(m)$ and $\Delta(m) \simeq \Dw + [2/(\beta
  L)]K(m)[H_0^2\sqrt{1+\zeta^2}-2\zeta K(m)]$. 
Eliminating $K(m)$ while keeping only the leading term in $\epsilon$,
we find the pressure-displacement relation,
\begin{equation}
  m\rightarrow 1:\ \ \ 
  P(\Delta) =\Pw - \frac{\alpha^2}{\beta(1+\zeta^2)}
  \left(\Delta-\Dw\right)^2
  \simeq \Pw - \frac{1}{16}(1 - 3\zeta/2)
  \left(\Delta-\Dw\right)^2.
\label{PDeltafold}
\end{equation}
This localized limit no longer depends on the system size
$L$. Equation (\ref{PDeltafold}) is to be compared with the result for
an infinite incompressible sheet, $P(\Delta) = 1 - \Delta^2/16$
\cite{Diamant2011}. We use this result to rewrite the localized
profile (\ref{localized}) in terms of the displacement,
\begin{eqnarray}
  m\rightarrow 1:\ \ \ h(s) &=& A_0 \frac{\cos(\kc s)}{\cosh(\eta s)}, \\
  A_0 &=& \left(\frac{2\alpha}{1+\zeta^2}\right)^{1/2} (\Delta-\Dw)
  \simeq \frac{\Delta-\Dw}{2} (1-5\zeta/8), \nonumber \\
  \eta &=& \frac{\alpha}{\sqrt{1+\zeta^2}} (\Delta-\Dw)
  \simeq \frac{\Delta-\Dw}{8} (1-5\zeta/4),
\end{eqnarray}
which, for $\zeta=0$, coincides with the infinite-length incompressible
case ($A_0=\Delta/2$, $\eta=\Delta/8$) \cite{Diamant2011}. Thus,
compressibility is found to lower and widen the fold.

Finally, we substitute $H(S)$ of Eq.~(\ref{Hfold}) in
Eq.~(\ref{gamma2}) to obtain the compression field in the folded
state,
\begin{equation}
  \gamma(s) = \gw + \epsilon^2 \zeta
  \left(1 - \frac{\kappa^2}{\alpha}
  \cos^2(\kc s) \dn^2(\kappa\epsilon s,m) \right)
  \xrightarrow{m\rightarrow 1} \gw + \half \zeta A_0^2
  \left( \frac{\alpha}{\beta} - \frac{\cos^2(\kc s)} 
  {\cosh^2(\eta s)} \right).
\end{equation}
The maximum compression is at the tip of the fold.

Figure~\ref{fig_pattern}(c) presents the height profile of a folded
sheet, along with the corresponding profile in the incompressible case
(dotted curve). In Fig.~\ref{fig_pattern}(d) we show the accompanying
localized compression field.

\section{Conclusion}
\label{sec_discuss}

\begin{figure}[tbh]
\vspace{0.7cm}
\centerline{\resizebox{0.48\textwidth}{!}
{\includegraphics{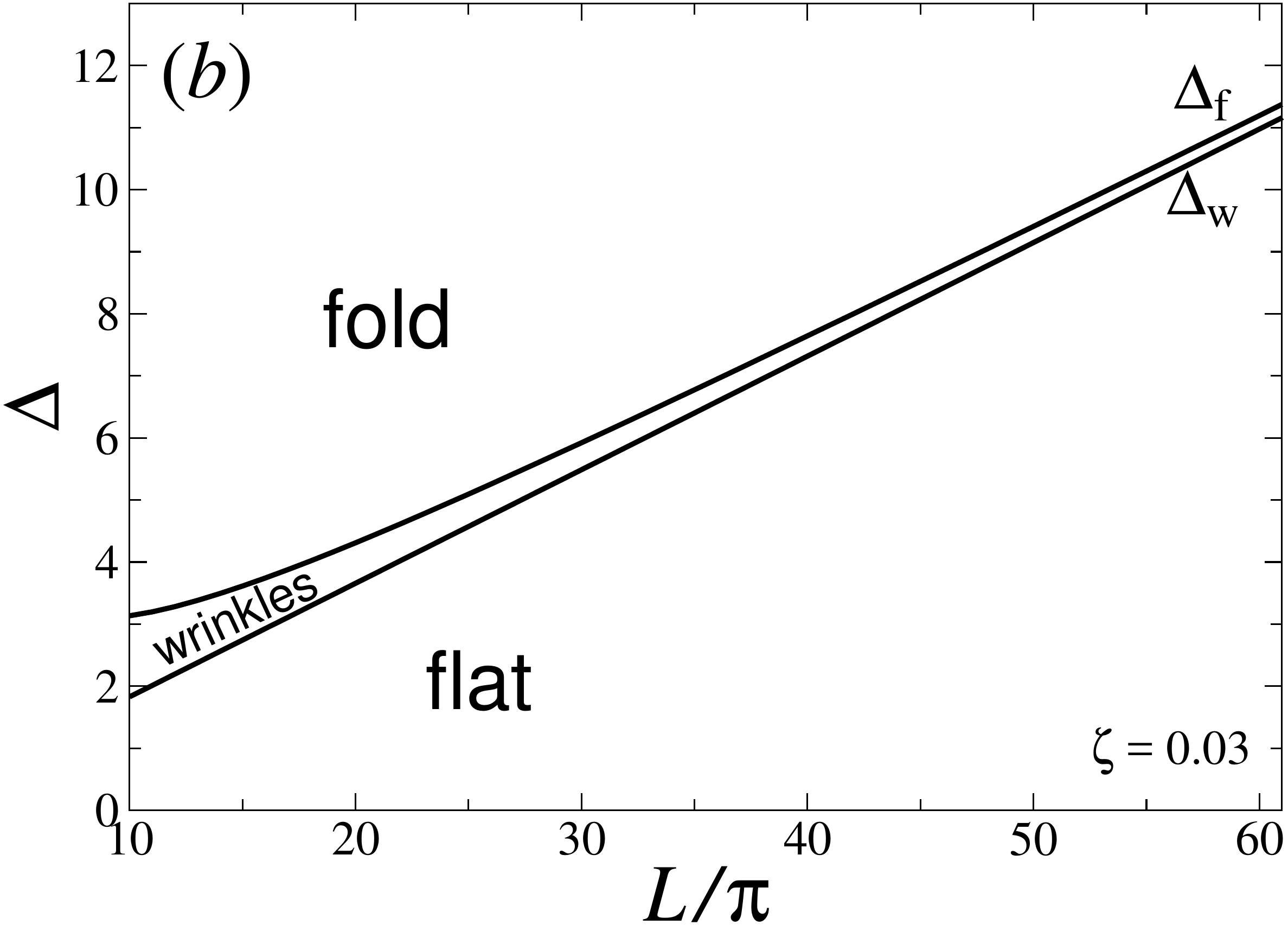}}
\hspace{0.2cm}
\resizebox{0.48\textwidth}{!}
{\includegraphics{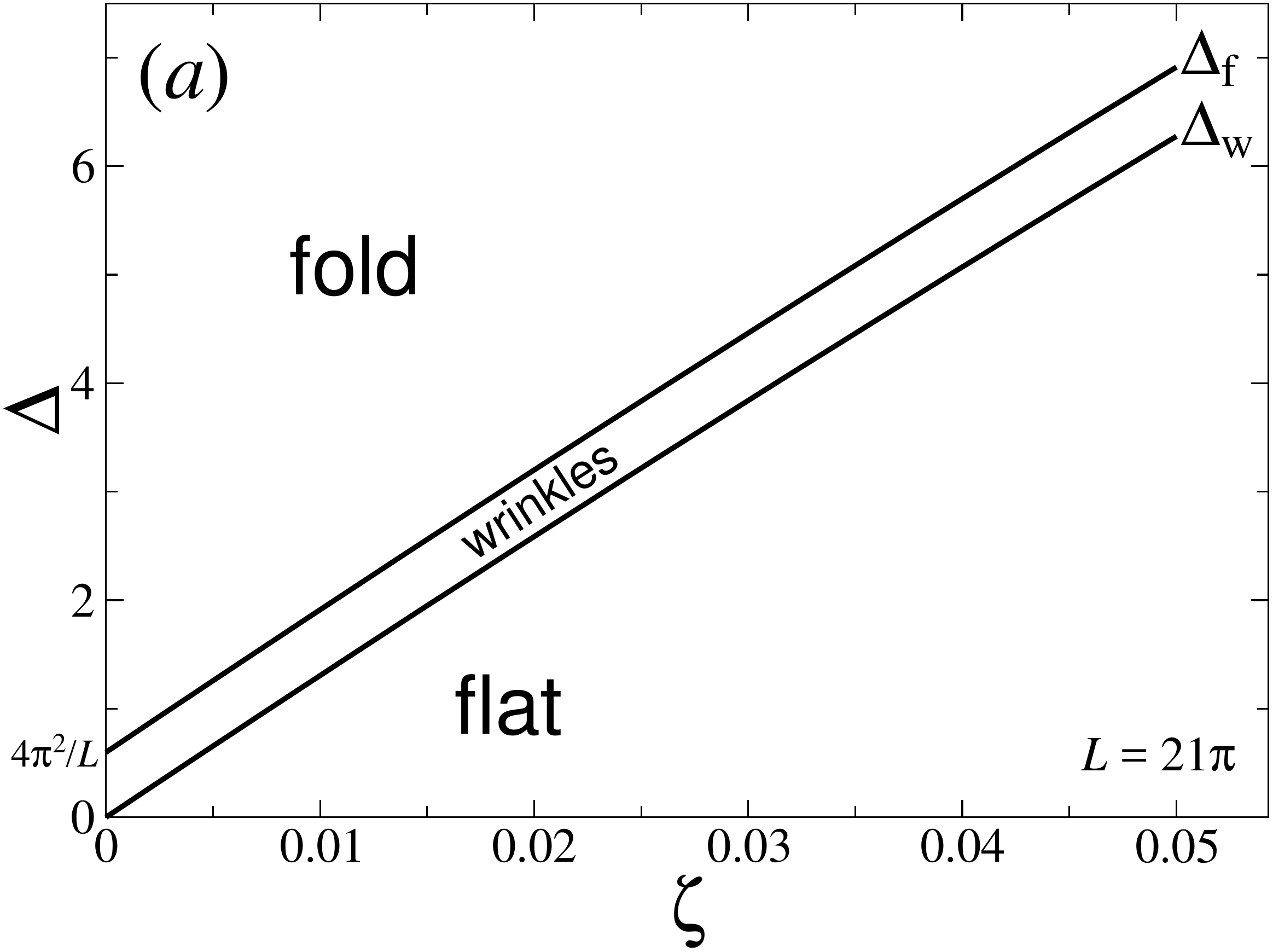}}}
\caption{State diagrams on the $(\Delta,\zeta)$ plane for fixed
  $L=21\pi$ (a) and $(\Delta,L)$ plane for fixed $\zeta=0.03$ (b). The
  displacement $\Delta$ and system length $L$ are rescaled by the
  intrinsic length $q^{-1}$ ($q$ being the wrinkle wavenumber of the
  incompressible sheet).}
\label{fig_diagram}
\end{figure}

In this work we have analyzed the effect of the compressibility of a
finite-size floating sheet on the patterns that the sheet develops
upon uniaxial lateral compression. The theory reproduces earlier
results for incompressible infinite \cite{Diamant2011} and finite
\cite{Oshri2015} floating sheets in the appropriate limits. It
provides the corrections introduced by finite compressibility. It
yields also the nonuniform compression fields (obviously absent in the
incompressible case), which accompany the buckling patterns.

The obtained information is concisely summarized in the two state
diagrams presented in Fig.~\ref{fig_diagram}. Figure
\ref{fig_diagram}(a) shows the regions of stability of the three
states\,---\,flat, wrinkled, and folded\,---\,on the
displacement--compressibility plane for a fixed sheet length. Figure
\ref{fig_diagram}(b) shows the stability regions of the same states on
the displacement--length plane for a fixed compressibility
parameter. The lines separating the stability regions represent
second-order transitions. In the limit of an incompressible sheet
($\zeta\rightarrow 0$) the flat-state region vanishes. The stability
region of the wrinkled state is narrow and disappears in the limit of
an infinite sheet ($L\rightarrow\infty$). Thus, diagrams containing
all three states require an account of both finite size and finite
compressibility, as we have presented here.

By expanding in $\epsilon^2=\Pw-P$ we have implicitly assumed that
$\zeta\gg\Pw-\Pf \simeq 8\pi^2/L^2$. This imposes an upper bound on
the sheet thickness $t$ (equivalently, a lower bound on its length
$L$) for the theory to be valid. In terms of the system's physical
parameters the condition reads
\begin{equation}
  t \ll \frac{\rho g L^2}{8\pi^2 E}.
\label{tcondition}
\end{equation}

Another practical consideration is that in typical experimental
scenarios $\zeta$ is very small. For example, the 10-$\mu$m thick
polyester sheet of Ref.~\cite{Pocivavsek2008}, the 100-nm thick
polystyrene sheets of Ref.~\cite{King2012}, and the lipid monolayers
of Refs.~\cite{Gopal2006a,Gopal2006b}, have $\zeta\sim 10^{-6}$,
$10^{-6}$, and $10^{-7}$, respectively. In these systems it would be
extremely difficult to measure the compressibility effects derived
here. On the other hand, there are softer sheets where these effects
may not be as minute. For example, the Young modulus of common
hydrogels is as low as 1--10~kPa. A 100-$\mu$m-thick layer of such a
gel will have $\zeta$ of a few percent. The validity condition
(\ref{tcondition}) for this system requires that $L>1$~cm. In fact,
elastic moduli of very soft materials are notoriously hard to measure
using conventional methods. The wrinkle wavelength is used to measure
the bending modulus of sheets, from which, if the thickness is known,
the Young modulus of the material can be indirectly inferred. One may
be able to resolve the delicate compressibility effects in soft
sheets, \eg by high-precision measurement of the wrinkle
wavelength. The wavelength derived above should be corrected for the
fact that, in practice, one observes the compressed, rather than
relaxed, lengths. Using Eqs.~(\ref{PW}), we get the apparent
wavelength as
\begin{equation}
  \lambda_{\rm apparent} = \gw\lambda_{\rm c} = \lambda \left(
  \sqrt{1+\zeta^2}-\zeta \right)^{1/2} \simeq \lambda(1-\zeta/2),
\end{equation}
where $\lambda=2\pi[B/(\rho g)]^{1/4}$ is the commonly used wrinkle
wavelength of an incompressible sheet. Thus, unlike the predicted
wavelength in the relaxed frame, the apparent one will be slightly
shorter than its incompressible counterpart. The correction will scale
as $t^{5/4}$, on top of the dominant $t^{3/4}$ term.  Such
measurements may provide another, independent and more direct, handle
for extracting small elastic moduli.

\acknowledgments

This work has been supported by the Israel Science Foundation (Grant
No.~164/14).

\appendix

\end{document}